\documentclass[10pt,conference]{IEEEtran}
\IEEEoverridecommandlockouts
\usepackage{cite}
\usepackage{amsmath,amssymb,amsfonts}
\usepackage{algorithmic}
\usepackage{graphicx}
\usepackage{textcomp}
\usepackage{xcolor}
\usepackage{siunitx}
\usepackage[caption=false]{subfig}
\def\BibTeX{{\rm B\kern-.05em{\sc i\kern-.025em b}\kern-.08em
    T\kern-.1667em\lower.7ex\hbox{E}\kern-.125emX}}
\begin{document}

\title{Symbol-Synchronous Communication for Ultra-Low-Power Multi-Hop Ambient IoT Networks

\thanks{This work was supported in part by the European Union’s Horizon Europe framework under Grant 101093046 - project OpenSwarm, in part by the Flemish Government under FWO Project G019722N - LOCUSTS, and in part by the AMBIENT-6G project, which received funding from the Smart Networks and Services Joint Undertaking (SNS JU) under the European Union’s Horizon Europe research and innovation program under Grant Ament No. 101192113. The computing resources and services for this work were supported by the HPC core facility CalcUA of the University of Antwerp, and Flemish Supercomputer Center~(VSC), funded by the Flemish government.}
}

\author{\IEEEauthorblockN{
                Xinlei Liu,
                Andrey Belogaev,
                and Jeroen Famaey}
\IEEEauthorblockA{IDLab, University of Antwerp -- imec, Belgium\\
        Email: \{Xinlei.Liu, Andrei.Belogaev, Jeroen.Famaey\}@uantwerpen.be}
}

\IEEEpubid{\begin{minipage}{\textwidth}\centering\ \\[12pt] 
  Copyright (c) 2026 IEEE. Personal use of this material is permitted.\\ 
  However, permission to use this material for any other purposes must be obtained from the IEEE by sending a request to pubs-permissions@ieee.org.
\end{minipage}} 

\maketitle
\begin{abstract}
Ambient Internet of Things (A-IoT) devices, as a critical enabler of future green IoT networks, have attracted broad interest from both industry and academia due to their ability to operate without batteries and with low maintenance costs. To accommodate their dynamic and constrained energy budget, an ultra-low-power connectivity protocol is required. Due to the severely limited transmit power of A-IoT devices, multi-hop connectivity is an interesting paradigm to extend their range. However, commonly used protocols for multi-hop communication may not be suitable for A-IoT due to excessive overhead related to channel access procedures, coordinated routing, and tight time synchronization requirements. This paper presents a novel network connectivity protocol based on symbol-synchronous transmissions, which allows battery-less relay nodes to participate in the forwarding process in an ad-hoc manner, without the need for synchronization or coordination. This allows them to adapt their duty cycle to the available harvested energy. Simulation results show that the proposed protocol can ensure high reliability in data packet delivery while significantly reducing the energy consumption of each relay node. We also investigate the relationship between wake-up probability and network density. For example, a $400$-node network in a $625\,\mathrm{m}^2$ area can achieve a packet error rate below \SI{1}{\percent} with an average awake time of \SI{6}{\percent} per node, achieving an energy consumption reduction of \SI{88}{\percent} compared to the baseline approach. 
\end{abstract}

\begin{IEEEkeywords}
Ambient IoT, symbol-synchronous communication, ultra-low-power, multi-hop, wireless sensor network~(WSN).
\end{IEEEkeywords}

\section{Introduction}
According to forecasts~\cite{gsma_report}, the number of Internet of Things (IoT) connections will exceed $40$ billion by 2030. Individual IoT networks can consist of hundreds or even thousands of connected devices, which are deployed in industrial factories, farms, buildings, forests, and other sites, depending on the application~\cite{jouhari2023survey}. Such large-scale deployments typically require long-distance coverage and low-power operation, as it is impossible to connect all IoT devices to a continuous power source. To address these issues, the A-IoT concept has been proposed, which considers maintenance-free devices powered by energy harvesting (EH), i.e., devices capable of converting ambient energy (e.g., solar, thermal, RF) into usable electrical energy for their electronic components~\cite{lopez2025zero}.
\IEEEpubidadjcol
To maintain connectivity for such systems, a suitable wireless communications technology is needed. Low Power Wide Area Networks (LPWANs) in a star topology are considered to be a promising solution due to their low-power profile. It has been shown that LPWANs can provide coverage in the range of several kilometers in open areas without impediments. However, current LPWAN solutions like LoRaWAN and NB-IoT have an average power consumption on the order of several \SI{100}{\milli\watt} during transmission~\cite{capuzzo2021ns, sultania2023batteryless}, which is often infeasible for low-power A-IoT devices that rely on ambient energy harvesting. Multi-hop networking is an alternative solution to extended coverage, without the need for high transmit power or complex modulation and coding schemes~\cite{islam2023performance}. Multi-hop networks allow devices to act as relays, thereby supporting communication between a distant initiator and sink via intermediary nodes. Although existing wireless multi-hop networks have proven to be a promising solution for many IoT applications~\cite{barrachina2017multi}, they require complex coordination protocols, like tight time synchronization, coordinated channel access, and network-wide routing, which makes them challenging to use with A-IoT devices.

Recently, symbol-synchronous communication has shown promise as a low-latency alternative to store-and-forward routing in multi-hop networks~\cite{oostvogels2020zero, liu2024low}. 
It allows multiple relay nodes to transmit the same signals concurrently as soon as a single symbol is detected. The main advantage of this approach is that it does not require additional procedures for time synchronization, coordinated routing, and channel access. While the concept was originally proposed to overcome the high latency introduced by such procedures, we hypothesize that this concept also makes it highly suitable for A-IoT networks, as the intermittent nature of A-IoT devices makes coordination and synchronization exceedingly hard.

Previously proposed symbol-synchronous transmission protocols assume relay nodes are always on, and participate in the forwarding of each symbol~\cite{oostvogels2020zero, liu2024low}. Keeping the radio in reception mode at all times is infeasible for A-IoT devices, which often need to sleep to recover their energy supply. Since symbol-synchronous transmission relies on constructive interference among concurrent transmissions of multiple relays for signal detection, the sink node can still correctly detect the transmitted symbol even if only a subset of relays forward each symbol, provided that the received signal strength is sufficiently high to be distinguished from noise. In other words, the network can achieve reliable delivery if the node density is sufficiently high. This property enables symbol-synchronous transmission to naturally support adaptive relay duty cycles, where each A-IoT relay forwards only a subset of symbols according to its available harvested energy. As a result, the overall energy consumption of the network can be dynamically adjusted without compromising the reliability of symbol delivery. To validate our hypothesis, we propose a simple randomized procedure in which the relay nodes sleep or wake up according to a fixed probability. Using a MATLAB-based simulation framework, we evaluate the network energy performance for varying network densities and wake-up probabilities.


The paper is structured as follows. In Section~\ref{sec:related_work}, we review the connectivity protocols used in A-IoT networks and analyze their limitations. In Section~\ref{sec:system_model}, we describe the protocol design comprising symbol-synchronous transmissions, modulation, detection, and a randomized wake-up procedure. In Section~\ref{sec:simulation}, we evaluate the performance of the proposed network in terms of reliability and energy consumption. Finally, in Section~\ref{sec:conclusion}, we outline the main conclusions along with the future work. 

\section{Related work}
\label{sec:related_work}
Wireless connectivity protocols for A-IoT networks currently rely on commercially mature network protocols, such as LPWANs (e.g., LoRaWAN or NB-IoT), Bluetooth Low Energy~(BLE), IEEE~802.15.4 varieties (e.g., ZigBee, 6TiSCH, or WirelessHART)~\cite{callebaut2021art, pereira2020challenges}. All of them allow communication at a low energy budget due to simple modulation schemes, the use of narrowband spectrum, low transmission power, and especially due to duty cycling, which keeps the devices inactive most of the time. Some of these technologies, e.g., LoRaWAN, operate in the sub-GHz band, thus providing extended coverage of up to several kilometers. However, the required power, especially peak power higher than \SI{100}{\milli\watt}, is not suitable for A-IoT devices with low EH efficiency. 

To guarantee reliable coverage and lower power consumption, multi-hop networks can be leveraged. In such networks, some of the nodes operate as relay nodes, which help the destination node receive data by forwarding the packets overheard from other nodes. Although multi-hop networks have been proven highly efficient in IoT deployments, their applicability is very limited when it comes to A-IoT networks. This is due to the fact that the need for accurate time synchronization and the coordination overhead among relays significantly limit the ability of A-IoT devices to act as relays. For example, the procedures of network joining and periodic synchronization, highlighted in~\cite{van2024integrating}, are too energy-intensive for intermittent A-IoT devices. To avoid packet collisions, routing procedures are required, which should account for situations when one or multiple nodes on the path have insufficient energy for data forwarding~\cite{poornima2023holistic}. Finally, conventional multi-hop networks use a store-and-forward routing approach, where the nodes forward the packet only after they receive all of its bits and verify the correct reception, usually using a checksum. Under this pattern, additional power is wasted due to listening to the channel, coordination, and routing.

As an alternative, the Low-power wireless bus (LWB)~\cite{ferrari2012low} was proposed to replace the standard IoT network stack. Specifically, routing is eliminated by leveraging concurrent transmissions among relay nodes. The basis of LWB is Glossy~\cite{ferrari2011efficient}, a protocol based on IEEE 802.15.4, which exploits constructive interference from concurrent transmissions of the same packets. However, Glossy imposes a very strict requirement on the time synchronization of \SI{0.5}{\micro\second} among the nodes, which is challenging for A-IoT networks.

Further building on the idea of concurrent transmissions, the concept of symbol-synchronous transmission, dubbed Zero-Wire, was proposed for wireless optical communications~\cite{oostvogels2020zero}. It breaks the paradigm of store-and-forward routing and demonstrates that the end-to-end latency can be further reduced to sub-millisecond levels by forwarding the information symbol by symbol. 
In Zero-Wire, the overlapping signals from multiple relays strengthen the resulting signal, as the interference in the optical domain is always constructive. Our own recent works~\cite{fang2024abl, liu2024low} explored the latency performance of Zero-Wire in the radio frequency~(RF) bands at 2.4 and 60 GHz. Simulation results showed that the Zero-Wire approach can be successfully applied to these bands, providing low latency, high scalability over multiple hops, and reliable data delivery. Given that symbol-synchronous transmission does not require tight time synchronization and coordination among relays, and relays can participate in the forwarding in an ad-hoc manner, it is naturally well-suited for intermittently powered A-IoT devices. This paper extends our prior research by investigating and optimizing the energy consumption of multi-hop networks using symbol-synchronous transmissions, which is crucial for the connectivity of A-IoT devices.

\section{System model}
\label{sec:system_model}
We consider a wireless multi-hop network operating in an RF band. Every node in the network can transmit data either periodically or triggered by some event, e.g., an abnormally high temperature or humidity rise. For each data packet, one node acts as the source and aims to deliver its packet to the sink node, which acts as a gateway (e.g., connected to an edge server). As the sink node is likely to be outside the direct communication range of the source node, multiple relay nodes cooperatively forward the data packet to the sink node based on the symbol-synchronous principle. All nodes in the network, except for the sink node, are considered low-power A-IoT devices relying on EH, and may not always have enough energy to be active. The sink node is assumed to have a stable power supply (e.g., a primary battery or connected to an external power supply), and it is always on.

\begin{figure}[!t]
\centering
\includegraphics[width=3.4in]{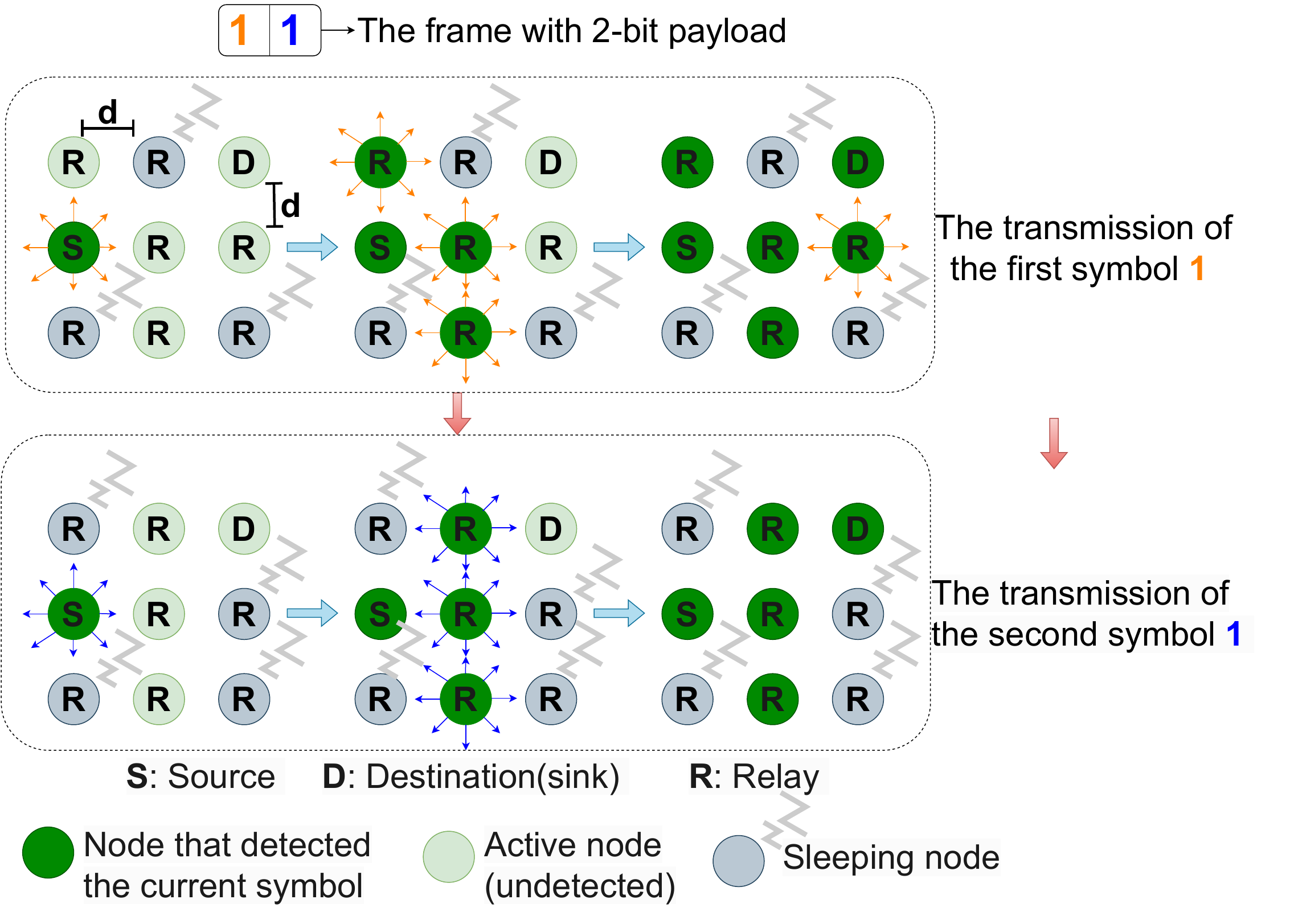}
\caption{System model}
\label{fig1}
\end{figure}

Fig.~\ref{fig1} shows a schematic example of the transmission of a small $2$-bit packet based on our system design. In Fig.~\ref{fig1}, the packet, consisting of $2$ $1$-bits, is transmitted and relayed symbol by symbol over multiple hops, which is the core idea of the symbol-synchronous transmission protocol~(cf., Section~\ref{section3_2}). As shown in the figure, relay nodes may individually decide to only forward a subset of bits, in order to save energy by going into sleep mode (cf., Section~\ref{section3_3}).

\subsection{Symbol-synchronous transmission}\label{section3_2}

The symbol-synchronous transmission paradigm is the core component of the system. It allows relay nodes to forward the information symbol by symbol. Specifically, every relay node immediately relays the symbol upon successful detection, without waiting for the reception of the whole packet and coordination with other nodes. This protocol exploits concurrent transmissions of multiple relay nodes, which means that multiple nodes are allowed to transmit the same symbol simultaneously. This is motivated by the fact that the interference from signal collisions is constructive under specific circumstances, which makes concurrent transmissions naturally advantageous, especially in denser deployments. Importantly, symbol-synchronous transmission does not require any complicated routing or collision avoidance procedures. In previous work, we have shown that even without explicit time synchronization across relays, the concurrent transmission of symbols still works~\cite{liu2024low}. These factors make the symbol-synchronous paradigm naturally suited for multi-hop communication with A-IoT devices, which, due to their limited and intermittent energy, have trouble maintaining synchronization, and cannot handle complex routing and channel access methods. Below, we briefly describe the modulation, relaying, and symbol detection procedures. For a more detailed description, the reader is referred to our previous work~\cite{liu2024low}. 

\begin{figure}[!t]
\centering
\includegraphics[width=3.4in]{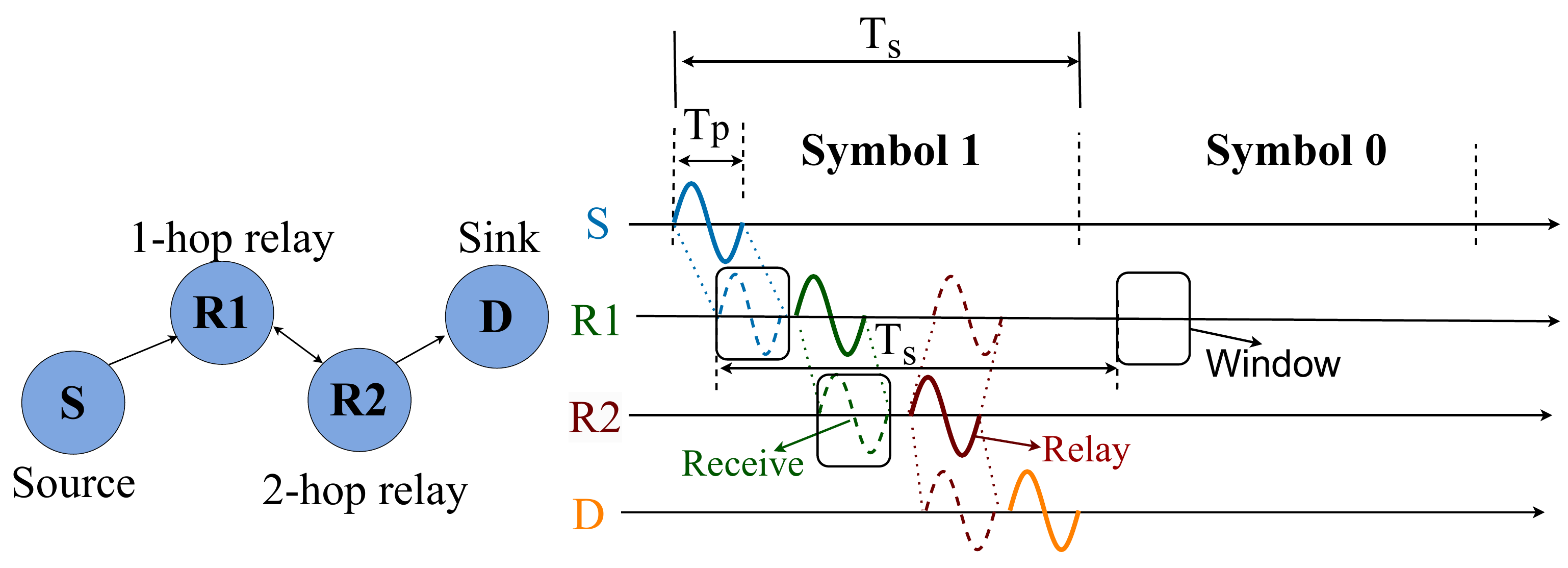}
\caption{Symbol-synchronous transmission}
\label{fig2}
\end{figure}

An on-off keying (OOK) modulation is often used for energy-constrained IoT devices due to its low complexity and high robustness even for very low transmission power~\cite{guo2025low}. Moreover, OOK modulation can more effectively demodulate non-synchronized, overlapping pulses from concurrent transmissions. Motivated by those, we apply short-pulse-based on-off keying~(POOK) modulation. As illustrated in Fig.~\ref{fig2}, for the symbol $1$, a short pulse with pulse duration $T_p$ is sent at the start of the symbol period $T_s \gg T_p$, followed by a silent guard to avoid inter-symbol interference (ISI). For the symbol $0$, the transmitter remains silent during the entire symbol period. 

Fig.~\ref{fig2} schematically shows how symbol-synchronous transmission is performed in a simple $4$-node network. The network consists of a source node, a sink node, and $2$ relay nodes, so that the data packet passes through $2$ hops. For the symbol $1$, the source node first sends the short pulse to the nearest relay node $R1$. After some propagation and processing delay, the node $R1$ receives the pulse and detects it. After that, $R1$ retransmits the POOK-modulated symbol to its neighboring node $R2$ and the source node. However, the source node operates only as a transmitter and therefore does not receive any signals. Then, relay node $R2$ follows the same procedure to detect and forward the symbol to the nodes close to $R2$, nodes $R1$ and $D$~(sink). In symbol-synchronous communication, the symbol period $T_s$ is required to be long enough so that all active nodes within mutual communication range manage to detect the symbol before the start of transmission of the next symbol, thus avoiding the relayed version of the current symbol from interfering with the detection of the next symbol. For symbol $0$, all transmitters remain silent, unless relay nodes accidentally misdetect symbol $0$ as symbol $1$ due to noise. As shown in Fig.~\ref{fig2}, the symbol period for each node starts when the symbol first arrives at that node, thus avoiding complicated synchronization with the source node. 

Similar to~\cite{liu2024low}, we use a window-based detector to receive POOK-modulated symbols. As shown in Fig.~\ref{fig2}, relay nodes are likely to receive multiple copies of the same signals in a single symbol period. These duplicates arrive at the node at different moments, causing the risk of detecting the same symbol multiple times. Additionally, repeated detection consumes more energy. Considering this problem, we apply a detection window whose duration is shorter than the symbol period to filter out these duplicates and save energy. Every node is only allowed to receive and detect signals in this window while remaining asleep outside of it. Once detection occurs, the window immediately slides to the start of the next symbol period. In every detection window, a voting method is utilized to make the decision. Specifically, symbol $1$ is detected if the majority of the samples' amplitudes exceed the predefined threshold. Otherwise, the symbol $0$ is detected. The threshold value is set to 9 dB higher than the noise floor. Note that the described procedure does not require time synchronization, and the position of the window may vary from node to node. Since this paper focuses mainly on the energy efficiency of symbol-synchronous transmission, more details about the detector logic can be found in our previous work~\cite{liu2024low}.
A preamble byte consisting of multiple (in this paper fixed to $8$) symbols $1$ is applied to resynchronize the window position of relay nodes and sink node at the start of every packet transmission. Also, the preamble byte allows the sink node to detect the start of a new packet transmission. The positions of initial detection windows for all nodes are defined once those nodes wake up and detect the symbol $1$. This long preamble offers sufficient observation opportunities for all nodes, thereby guaranteeing a reliable window positioning and detection of a new packet transmission. 

\subsection{Relay operation workflow}\label{section3_3}

For A-IoT deployments, it is not efficient to make all of the relay nodes participate in every symbol transmission, as A-IoT devices may not be able to harvest enough energy to forward all symbols of a packet. Therefore, we assume that every relay node in the network is configured to randomly wake up with a probability $P$ during each symbol period to participate in relaying the symbol. In other words, every relay node remains in a sleep state with probability $1 - P$ for energy saving and potential EH. In this initial study, the wake-up probability $P$ is assumed to be identical for every relay node. In future work, we will study how to dynamically adapt the probability $P$, based on the remaining energy of the node and EH conditions. Note that the sink node remains active for every transmission. 

\begin{figure}[!t]
\centering
\includegraphics[width=3.4in]{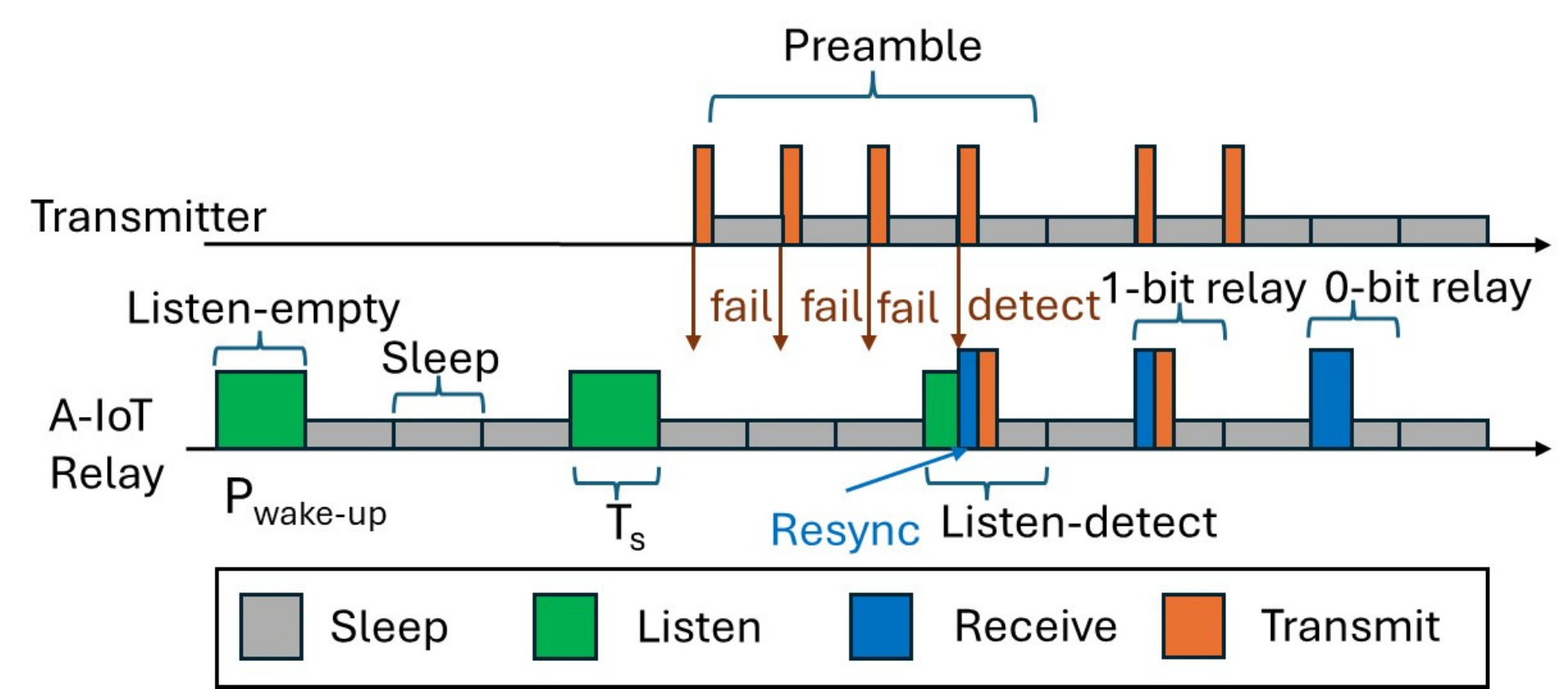}
\caption{States of relay node during: (1) listening and failing to detect the start of a new packet (listen-empty); (2) sleeping to save energy (sleep); (3) listening and managing to detect a new packet (listen-detect); (4) relaying of symbol $1$ ($1$-bit relay); (5) relaying of symbol $0$ ($0$-bit relay).}
\label{fig3}
\end{figure}

Fig.~\ref{fig3} depicts different states of a relay before and during the packet transmission. As there is no packet transmission schedule and no time synchronization, relay nodes do not have information on when a new packet transmission starts. To save energy, relay nodes do not listen all the time. Instead, they wake up with probability $P$ and listen during a single symbol period $T_s$. Once symbol $1$ is detected, the relay resynchronizes its detection window to the time the symbol was detected (shown in the figure as ``resync''). After that, this relay starts participating in the forwarding of each symbol with probability $P$. When symbol $1$ is relayed, the relay remains in the listening state until the successful detection, and then switches to transmission state to forward the pulse. Once forwarding is finished, it switches to the sleep state. When the symbol $0$ is relayed, the relay remains in the listening state for the entire duration of the detection window. After that, it directly switches to sleep.

\section{Performance evaluation}
\label{sec:simulation}
To evaluate the performance of the system, we conduct experiments using MATLAB. We study how the reliability of packet delivery depends on wake-up probability and network density, and estimate the energy consumption of the relays.

\subsection{Simulation setup}
In this paper, we simulate a wireless sensor network where the nodes are arranged in a square grid with a fixed grid distance $d$ (i.e., the distance between horizontal and vertical neighbors), as shown in Fig.~\ref{fig1}. The sink node is located in the upper-right corner and is fixed. The source node is selected randomly and varies from packet to packet. In addition, the TGac channel model~\cite{ieee2009tgac} is selected to simulate the signal propagation, which provides a realistic and sophisticated representation of multipath and delay-spread characteristics in practical environments. 
The transmission power of every node is fixed at \SI{0}{dBm}, which is a typical value for low-power IoT devices~\cite{yang20190}. The selected carrier frequency corresponds to the last channel of the \SI{2.4}{\giga\hertz}
ISM band, \mbox{2483.5-\SI{2500}{\mega\hertz}}. To account for hardware impairments, we add a random carrier frequency offset~(CFO) in the range of $\pm10\,\mathrm{kHz}$, which is inevitably introduced by an imperfect local oscillator~\cite{baddeley2023understanding}. We limit the channel bandwidth $B$ to around \SI{3}{\mega\hertz}, which is consistent with the well-established wireless standards in multi-hop sensor network applications, such as IEEE 802.15.4~\cite{pereira2020challenges}. A raised-cosine filter with a roll-off factor \( \alpha = 0.5 \) is applied to limit the signal bandwidth to \SI{3}{\mega\hertz}. The pulse duration before filtering can be derived via \( \frac{1 + \alpha}{B} = 0.5~\mu\text{s} \). Accordingly, the pulse duration $T_p$ after filter is \SI{3}{\micro\second}. Given the $T_p$ of \SI{3}{\micro\second}, the sampling rate is set to \SI{20}{\mega\hertz} to guarantee a sufficient number of received samples for voting. The window length is set to \SI{3}{\micro\second}. To meet the requirement of $T_p \ll T_s$ in the symbol-synchronous protocol, $T_s$ is selected to be \SI{25}{\micro\second}~(data rate is $1/T_s = 40$ kbps). Our experiments showed that this value ensures ISI does not occur. In order to estimate the energy consumption, we use the power draw values of a commercial IEEE 802.15.4-compliant Chipcon CC2538 transceiver~\cite{daneels2018accurate}, whose transmit power and bandwidth are comparable to those of our design. All significant parameters used in the simulation are listed in Table~\ref{table1}. It should be noted that, on practical deployments, network topologies and transmission power may deviate from the above assumptions due to the dynamic and complicated IoT environment. However, the proposed protocol remains valid under irregular multi-hop networks. In addition, the protocol leverages the constructive interference generated from concurrent transmission to recover the signals. Therefore, moderate variations in the transmission power of individual nodes do not affect the symbol detection. 

\begin{table}[!t]
\caption{List of parameters used in simulation\label{table1}}
\centering
\begin{tabular}{c|c}
\hline
\textbf{Parameter} & \textbf{Value}\\
\hline
Transmission power & \SI{0}{dBm}\\
The center frequency of the carrier & \SI{2491}{\mega\hertz}\\
CFO range & $\pm10\,\mathrm{kHz}$\\
Bandwidth & \SI{3}{\mega\hertz}\\
Roll-off factor of the raised-cosine filter & $0.5$\\
Pulse duration $T_p$ & \SI{3}{\micro\second}\\
Window length & \SI{10}{\micro\second}\\
Sampling rate & \SI{20}{\mega\hertz}\\
Data rate & \SI{40}{kbps}\\
Noise floor & \SI{-60}{dBm}\\
Power consumption when transmitting & \SI{94.41}   {\milli\watt}\\
Power consumption when receiving & \SI{80.82}{\milli\watt}\\
Power consumption in sleep state  & \SI{1.8}{\milli\watt}\\
\hline
\end{tabular}
\end{table}

We consider three grid-patterned wireless sensor networks with different node densities deployed in a fixed area of $625\,\mathrm{m}^2$~($25\,\mathrm{m} \times 25\,\mathrm{m}$). Specifically, \{$25$, $100$, $400$\}~nodes are deployed with a grid distance $d$ of \{$5$, $2.5$, $1.25$\}~meters, respectively. At the same time, a series of different wake-up probabilities~$P$ is considered to evaluate the energy consumption reduction that can be achieved while providing the required network reliability for different network densities. For every network with fixed $P$ and node density, $20000$ 128-bit packets are transmitted. In this paper, the packet error rate~(PER) and preamble loss rate~(PrLR) are adopted to indicate network reliability. To reduce the number of packets with bit errors, we apply error-correcting Bose-Chaudhuri-Hocquenghem (BCH) code~\cite{forney2003decoding} with a code rate of \SI{80}{\percent}. If the sink node fails to detect the start of a new packet transmission, we consider the transmitted packet lost and calculate the PrLR. For each experiment, we also calculate the average energy consumption of the relay nodes spent during the transmission of a \mbox{$128$-bit} packet.

\begin{figure*}[!ht]
\centering
\subfloat[$25$ nodes\label{re_25}]{%
    \includegraphics[width=2.3in]{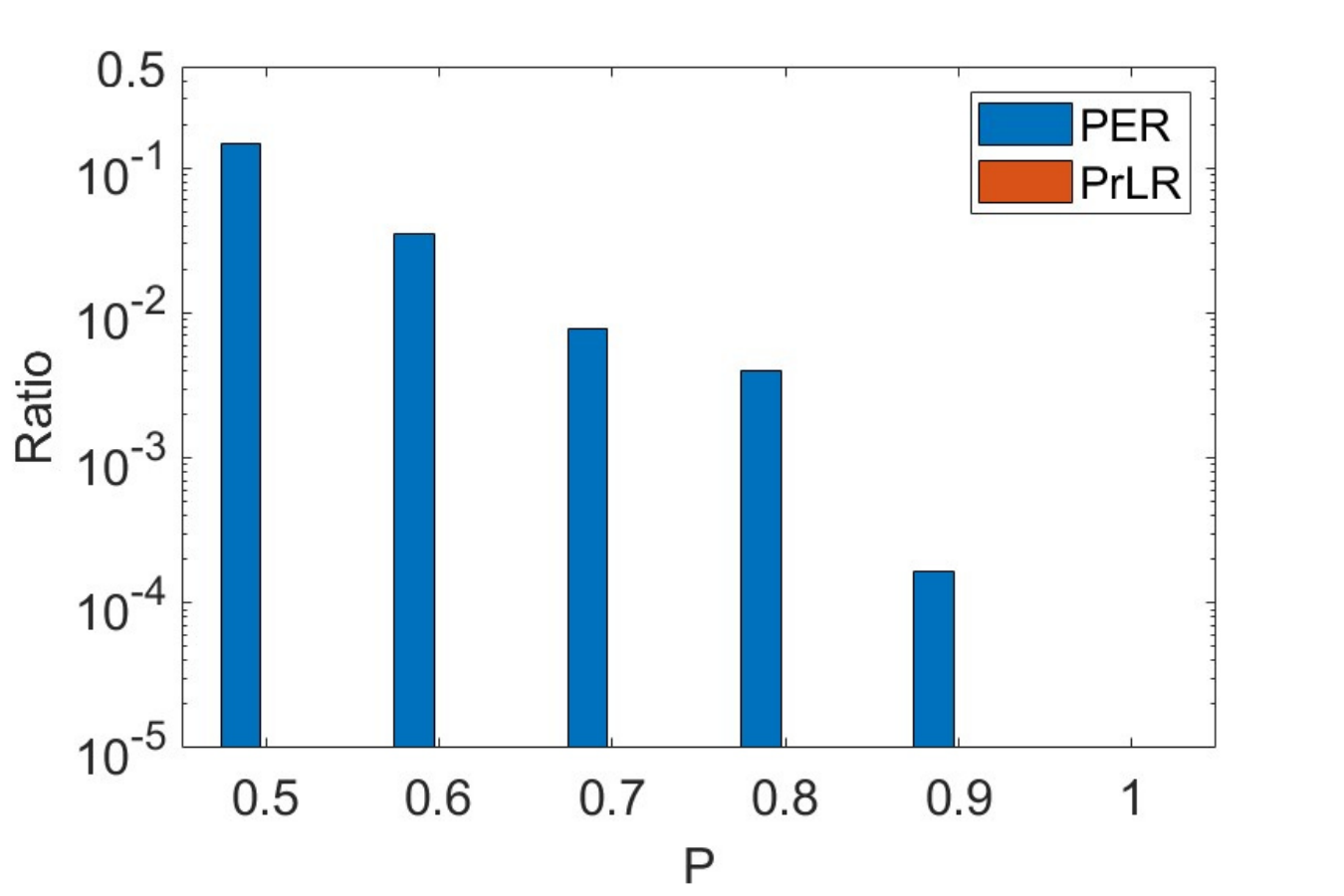}}
\hfil
\subfloat[$100$ nodes\label{re_100}]{%
    \includegraphics[width=2.3in]{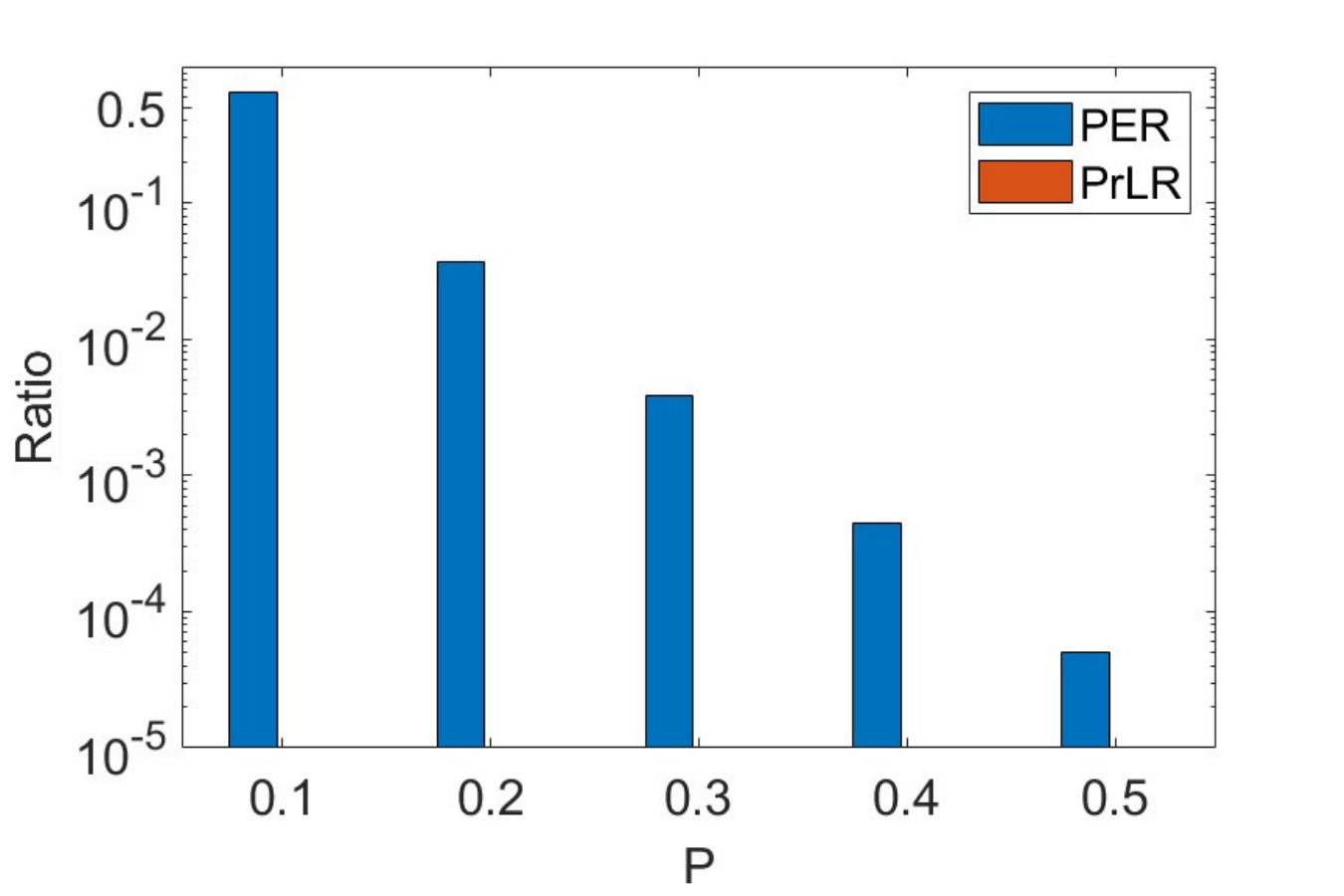}}
\hfil
\subfloat[$400$ nodes\label{re_400}]{%
    \includegraphics[width=2.3in]{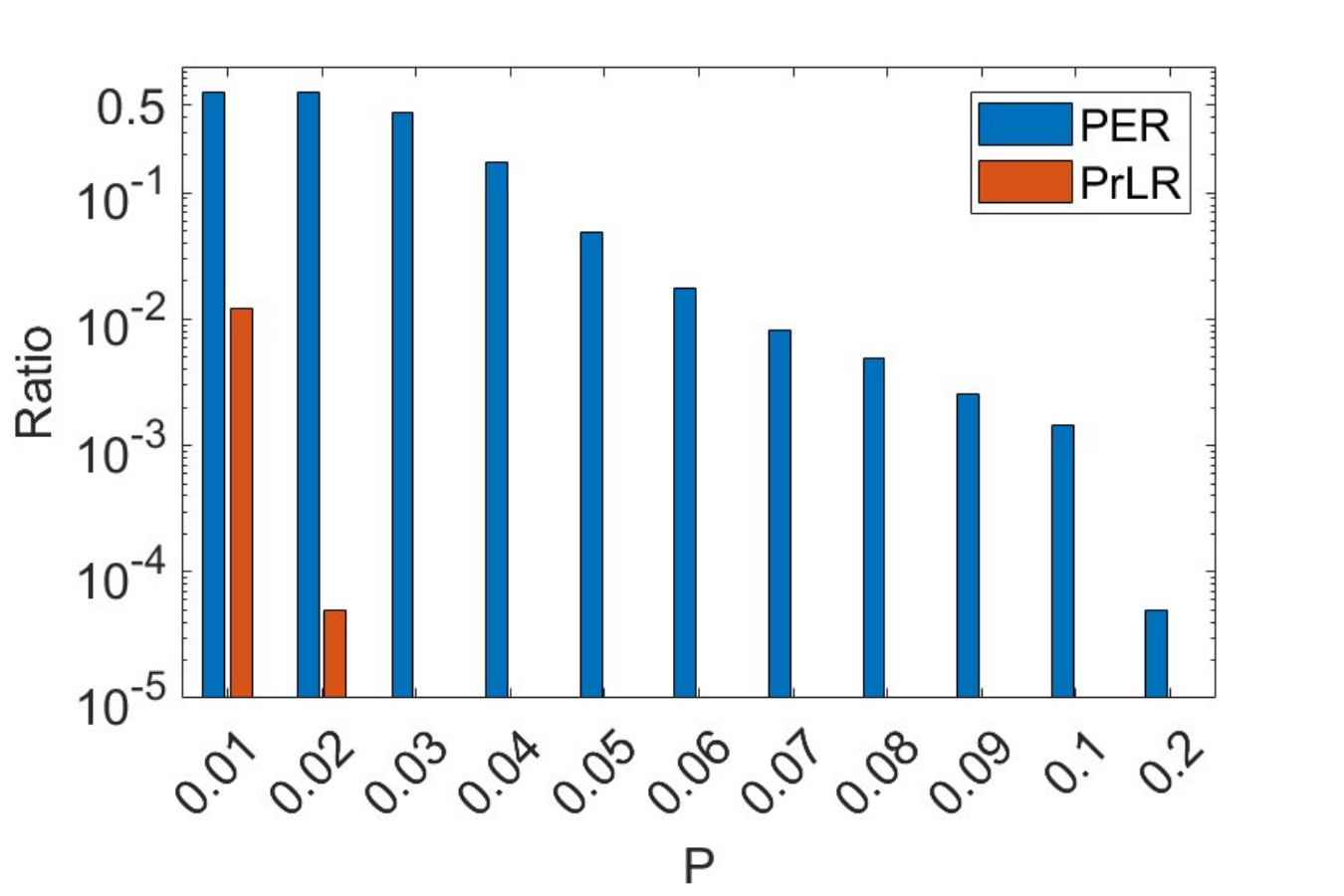}}
\caption{Network reliability, expressed as PER, as a function of wake-up probability for different network densities}
\label{fig4}
\end{figure*}


\begin{figure}[!t]
\centering
\includegraphics[width=3.4in]{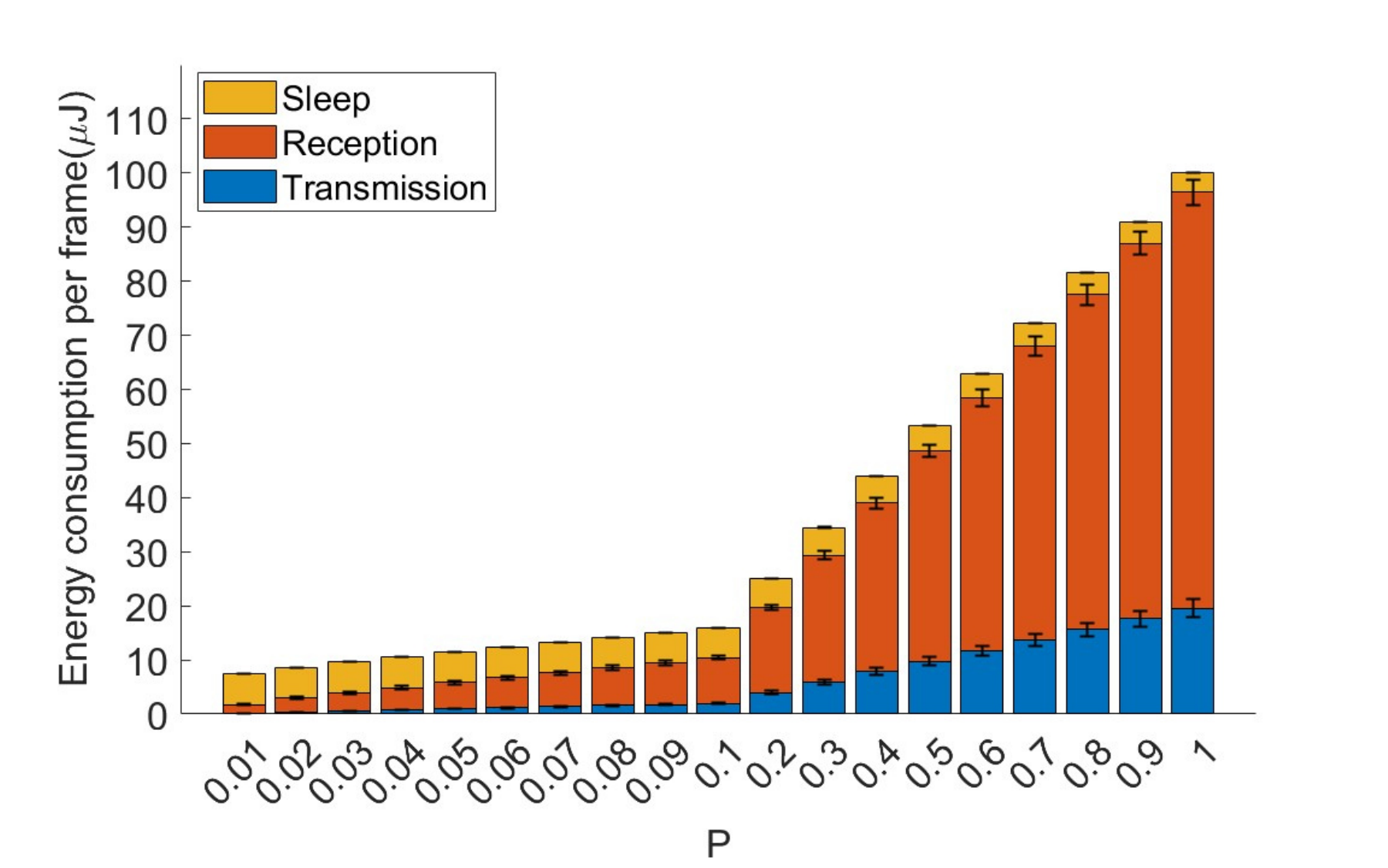}
\caption{The average energy consumption for transmitting one packet of $128$ bits}
\label{fig5}
\end{figure}

\subsection{Analysis of simulation results}

Fig.~\ref{fig4} shows the network reliability for different node densities as a function of wake-up probability. For a sparse \mbox{$25$-node} network~(Fig.~\ref{re_25}), the PER gradually decreases with the increase of wake-up probability $P$, reaching values below $10^{-2}$ when $P > 0.6$. Similarly, denser \mbox{$100$-node} and \mbox{$400$-node} networks show a similar descending trend as $P$ rises. However, with higher density, the network can achieve a similar PER with a smaller wake-up probability. Specifically, for the \mbox{$100$-node} network, only $P$ higher than $0.2$ is needed to reach the $PER < 10^{-2}$. For the \mbox{$400$-node} network, the lowest required $P$ is further decreased to $0.06$. This can be attributed to the fact that with higher density, more relay nodes participate in symbol forwarding, thereby increasing the probability of successful reception of the packet by the sink node. Similarly, under the same node density, a higher wake-up probability $P$ increases the number of active relay nodes in the symbol period, further boosting cooperative forwarding. Conversely, a very small $P$ like $0.01$ in the 400-node network sometimes causes failed detection of the preamble at the sink node, resulting in the packet loss~(Fig.~\ref{re_400}). 

We also evaluate the average energy consumption of a relay node for various wake-up probabilities for the transmission of a single 128-bit packet. Fig.~\ref{fig5} respectively shows the energy consumption in the transmission, reception, and sleep state. As illustrated in Fig.~\ref{fig5}, the transmission and reception energy increase with $P$, while the energy consumption of the sleep status decreases as the $P$ increases. The underlying reason is that, as the wake-up probability rises, the relay node spends more time in transmission and reception mode, while its sleep duration is reduced. Jointly analyzing Figs.~\ref{fig4} and~\ref{fig5}, it can be derived that, to maintain reliable packet delivery~($PER < 10^{-2}$) over the multi-hop network, every relay node in a sparse network~($25$ nodes) consumes at least \SI{62}{\micro\joule} per 128-bit packet. However, in a denser network, such as those with $100$ nodes and $400$ nodes, the corresponding energy consumption per 128-bit packet is reduced to \SI{25}{\micro\joule} and \SI{12}{\micro\joule}. 

We also evaluate the power consumption of a relay node in different states shown in Fig.~\ref{fig3}.
Table~\ref{table2} lists the average required power for each of the five states and the corresponding average energy consumption in a single symbol period. For the state \textit{listen-detect}, we mathematically derive the mean power consumption assuming that the moment when the node detects the new packet is uniformly distributed over the whole symbol period. As shown in Table~\ref{table2}, the $1$-bit and $0$-bit relays have similar average power consumption of around \SI{30}{\milli\watt}. The average power consumption for the states of $0$-bit relay, $1$-bit relay, and listen-empty is deterministic. The highest power consumption of \SI{80.83}{\milli\watt} is seen in the state of \textit{listen-empty}, since the node has to keep listening in the whole symbol period. Even in the worst-case state (listen-empty), the maximum power consumption of \SI{80.83}{\milli\watt} is still below the average power consumption of several \SI{100}{\milli\watt} in current LPWAN networks~\cite{capuzzo2021ns, sultania2023batteryless}. 


\begin{table}[!t]
\caption{Power consumption of a relay node in various states\label{table2}}
\centering
\begin{tabular}{c|c|c}
\hline
\textbf{State} & \textbf{Power} & \textbf{Energy per symbol period}\\
\hline
$1$-bit relay & $36.54 \pm 11.18~\mathrm{mW}$ & $0.91 \pm 0.28~\mu\mathrm{J}$
\\ 
$0$-bit relay & \SI{33.41}{\milli\watt} & \SI{0.84}{\micro\joule}\\
Sleep &  \SI{1.80}{\milli\watt} & \SI{45}{\nano\joule}\\
Listen-empty & \SI{80.82}{\milli\watt} & \SI{2}{\micro\joule}\\
Listen-detect & $52.14 \pm 17.38~\mathrm{mW}$ & $1.30 \pm 0.43~\mu\mathrm{J}$ \\
\hline
\end{tabular}
\end{table}

\section{Conclusion and Outlook}
\label{sec:conclusion}
In this paper, we analyze and optimize the energy consumption of a symbol-synchronous transmission protocol to enable energy-constrained A-IoT devices to operate as relays in wireless multi-hop networks. We investigate how to adapt the proposed protocol to support low-power A-IoT devices, allowing them to participate in the forwarding of only some of the symbols, while others can remain in sleep mode to save and/or harvest energy. This allows A-IoT devices to participate in the forwarding process as relays. This is generally not the case in classical multi-hop store-and-forward protocols, due to their need for coordinated listening and channel access, as well as time synchronization. Our simulation results show the trade-off between the optimal wake-up probability and node density. For example, a 400-node network deployed over $625\,\mathrm{m}^2$ can achieve reliable transmission (i.e., $PER<10^{-2}$) with a wake-up probability of just over $0.06$ and corresponding energy consumption of \SI{12}{\micro\joule} per $128$-bit packet, which saves around \SI{88}{\percent} energy compared to a node that is continuously on~\cite{oostvogels2020zero, liu2024low}. 

Our future work will focus on a more realistic wake-up strategy, which will dynamically adjust the wake-up probability $P$ according to the available energy at each node. In addition, we will also consider the actual, dynamic EH process on nodes. Moreover, a more complicated modulation scheme will be considered to improve the network data rate.

\bibliographystyle{IEEEtran}
\bibliography{references}

@article{guo2025low,
  title={{Low RF Power Harvesting Enabled Wireless Sensor Node With Long-Distance Communication Capability}},
  author={Guo, Lei and others},
  journal={IEEE Transactions on Microwave Theory and Techniques},
  year={2025},
  publisher={IEEE}
}

@misc{ieee2009tgac,
  author       = {{IEEE 802.11 Working Group}},
  title        = {{TGac Channel Model Addendum Document}},
  note         = {Document 11-09-0308-03-00ac},
  year         = {2009},
  month        = mar,
}

@article{yang20190,
  title={{A 0.2-V Energy-Harvesting BLE Transmitter with a Micropower Manager Achieving 25\% System Efficiency at 0-dBm Output and 5.2-nW Sleep Power in 28-nm CMOS}},
  author={Yang, Shiheng and others},
  journal={IEEE Journal of Solid-State Circuits},
  volume={54},
  number={5},
  pages={1351--1362},
  year={2019},
  publisher={IEEE}
}

@techreport{gsma_report,
    organization = "GSMA Intelligence",
    title = {{IoT market forecast to 2030: connections by region and vertical}},
    url = "https://www.gsmaintelligence.com/research/iot-market-forecast-to-2030-connections-by-region-and-vertical",
    year = 2024
}

@article{jouhari2023survey,
  title={{A Survey on Scalable LoRaWAN for Massive IoT: Recent Advances, Potentials, and Challenges}},
  author={Jouhari, Mohammed and others},
  journal={IEEE Communications Surveys \& Tutorials},
  volume={25},
  number={3},
  pages={1841--1876},
  year={2023},
  publisher={IEEE}
}

@article{lopez2025zero,
  title={{Zero-energy devices for 6G: Technical enablers at a glance}},
  author={L{\'o}pez, Onel and others},
  journal={IEEE IoT Magazine},
  volume={8},
  number={3},
  pages={14--22},
  year={2025},
  publisher={IEEE}
}

@article{islam2023performance,
  title={{Performance Evaluation of Multi-Hop LoRaWAN}},
  author={Islam, Md Rakibul and others},
  journal={IEEE Access},
  volume={11},
  pages={50929--50945},
  year={2023},
  publisher={IEEE}
}

@article{barrachina2017multi,
  title={{Multi-Hop Communication in the Uplink for LPWANs}},
  author={Barrachina-Mu{\~n}oz, Sergio and others},
  journal={Computer Networks},
  volume={123},
  pages={153--168},
  year={2017},
  publisher={Elsevier}
}

@inproceedings{ferrari2011efficient,
  title={Efficient network flooding and time synchronization with glossy},
  author={Ferrari, Federico and Zimmerling, Marco and Thiele, Lothar and Saukh, Olga},
  booktitle={Proceedings of the 10th ACM/IEEE International Conference on Information Processing in Sensor Networks},
  pages={73--84},
  year={2011},
  organization={IEEE}
}

@inproceedings{oostvogels2020zero,
  title={{Zero-Wire: a Deterministic and Low-Latency Wireless Bus through Symbol-Synchronous Transmission of Optical Signals}},
  author={Oostvogels, Jonathan and others},
  booktitle={Proc. of ACM SenSys},
  pages={164--178},
  year={2020}
}

@inproceedings{fang2024abl,
  title={{ABL: Leveraging Millimeter Wave Pulses for Low Latency IoT Networking}},
  author={Fang, Bingwu and others},
  booktitle={Proc. of CrystalFreeIoT},
  year={2024}
}

@inproceedings{liu2024low,
  title={{Low-latency Symbol-Synchronous Communication for Multi-hop Sensor Networks}},
  author={Liu, Xinlei and others},
  booktitle={Proc. of EuCNC},
  pages={1096--1101},
  year={2024}
}

@article{daneels2018accurate,
  title={{Accurate Energy Consumption Modeling of IEEE 802.15.4e TSCH using Dual-Band OpenMote Hardware}},
  author={Daneels, Glenn and others},
  journal={Sensors},
  volume={18},
  number={2},
  pages={437},
  year={2018},
  publisher={MDPI}
}

@article{callebaut2021art,
  title={{The Art of Designing Remote IoT Devices—Technologies and Strategies for a Long Battery Life}},
  author={Callebaut, Gilles and others},
  journal={Sensors},
  volume={21},
  number={3},
  pages={913},
  year={2021},
  publisher={MDPI}
}

@article{pereira2020challenges,
  title={{Challenges in resource-constrained IoT devices: Energy and communication as critical success factors for future IoT deployment}},
  author={Pereira, Felisberto and others},
  journal={Sensors},
  volume={20},
  number={22},
  pages={6420},
  year={2020},
  publisher={MDPI}
}

@article{van2024integrating,
  title={{Integrating Battery-Less Energy Harvesting Devices in Multi-Hop Industrial Wireless Sensor Networks}},
  author={Van Leemput, Dries and others},
  journal={IEEE Communications Magazine},
  volume={62},
  number={7},
  pages={66--73},
  year={2024},
  publisher={IEEE}
}

@article{poornima2023holistic,
  title={{Holistic Survey on Energy Aware Routing Techniques for IoT Applications}},
  author={Poornima, MR and others},
  journal={Journal of Network and Computer Applications},
  volume={213},
  pages={103584},
  year={2023},
  publisher={Elsevier}
}

@article{baddeley2023understanding,
  title={Understanding concurrent transmissions: The impact of carrier frequency offset and RF interference on physical layer performance},
  author={Baddeley, Michael and Boano, Carlo Alberto and Escobar-Molero, Antonio and Liu, Ye and Ma, Xiaoyuan and Marot, Victor and Raza, Usman and R{\"o}mer, Kay and Schuss, Markus and Stanoev, Aleksandar},
  journal={ACM Transactions on Sensor Networks},
  volume={20},
  number={1},
  pages={1--39},
  year={2023},
  publisher={ACM New York, NY}
}

@article{forney2003decoding,
  title={On decoding BCH codes},
  author={Forney, George},
  journal={IEEE Transactions on information theory},
  volume={11},
  number={4},
  pages={549--557},
  year={2003},
  publisher={IEEE}
}

@article{sultania2023batteryless,
  title={Batteryless NB-IoT prototype for bidirectional communication powered by ambient light},
  author={Sultania, Ashish Kumar and Famaey, Jeroen},
  journal={Ad Hoc Networks},
  volume={142},
  pages={103100},
  year={2023},
  publisher={Elsevier}
}

@inproceedings{capuzzo2021ns,
  title={An ns-3 implementation of a battery-less node for energy-harvesting internet of things},
  author={Capuzzo, Martina and Delgado, Carmen and Famaey, Jeroen and Zanella, Andrea},
  booktitle={Proceedings of the 2021 Workshop on Ns-3},
  pages={57--64},
  year={2021}
}

@inproceedings{ferrari2012low,
  title={Low-power wireless bus},
  author={Ferrari, Federico and Zimmerling, Marco and Mottola, Luca and Thiele, Lothar},
  booktitle={Proceedings of the 10th ACM Conference on Embedded Network Sensor Systems},
  pages={1--14},
  year={2012}
}

\end{document}